\documentclass[epj]{svjour}
\usepackage{graphics,latexsym,amssymb}
\usepackage{cite}

\begin{document}
\title{Finite-Size Scaling of the Level Compressibility at the\\
  Anderson Transition}
\author{M. L. Ndawana\inst{1}\thanks{\emph{Permanent
      address:} University of Zambia, Physics Department, Lusaka,
    Zambia} \and R. A. R\"{o}mer\inst{1}\thanks{\email{r.roemer@physik.tu-chemnitz.de}} \and M.
  Schreiber\inst{1,2}}
\institute{Institut f\"{u}r Physik Technische Universit\"{a}t, D-09107
  Chemnitz, Germany \and School of Engineering and Science, International University Bremen,
D-28725 Bremen, Germany}
\date{Received: date / Revised version: date / \mbox{$Revision: 1.39 $}}
\abstract{We compute the number level variance $\Sigma_{2}$ and the
  level compressibility $\chi$ from high precision data for the Anderson
  model of localization and show that they can be used in order to
  estimate the critical properties at the metal-insulator transition by
  means of finite-size scaling.  With $N$, $W$, and $L$ denoting,
  respectively, linear system size, disorder strength, and the average number
  of levels in units of the mean level spacing, we find that both
  $\chi(N,W)$ and the integrated $\Sigma_{2}$ obey finite-size scaling.
  The high precision data was obtained for an anisotropic
  three-dimensional Anderson model with disorder given by a box
  distribution of width $W/2$.  We compute the critical exponent as $\nu
  \approx 1.45 \pm 0.12$ and the critical disorder as $W_{\rm c} \approx
  8.59 \pm 0.05$ in agreement with previous transfer-matrix studies in
  the anisotropic model. Furthermore, we find $\chi\approx 0.28 \pm
  0.06$ at the metal-insulator transition in very close agreement with
  previous results.
 % end of PACS codes
} %end of abstract
\maketitle
%

%%%%%%%%%%%%%%%%%%%%%%%%%%%%%%%%%%%%%%%%%%%%%%%%%%%%%%%%%%%%%%%%%%%%%%%%%%
\section{Introduction}
\label{eq-sec-intro}
%%%%%%%%%%%%%%%%%%%%%%%%%%%%%%%%%%%%%%%%%%%%%%%%%%%%%%%%%%%%%%%%%%%%%%%%%%

The Anderson metal-insulator transition (MIT) in disordered systems
has been vigorously studied for a long time \cite{KraM93,LeeR85,And58}
and still continues to attract much attention \cite{Sch99a}.
For non-interacting electrons in disordered systems the scaling
hypothesis of localization has been successfully validated by
theoretical \cite{VolW82,VolW92} and numerical
\cite{PicS81a,PicS81b,MacK81,MacK83} approaches. The latter approaches
use well-known techniques of finite-size scaling (FSS) \cite{Bin97}.
FSS at the Anderson MIT has a noteworthy history, reaching a first
peak with the seminal papers of Pichard/Sarma \cite{PicS81a,PicS81b}
and MacKinnon/Kramer \cite{MacK81,MacK83}.  Especially in Ref.\
\cite{MacK83}, the groundwork for a reliable, numerical FSS procedure
was laid and scaling curves could be constructed that proved the
existence of an MIT in 3D and the absence of such in 2D and 1D. In these and
later studies based on the same analysis technique \cite{KraM93}, the
critical exponent $\nu$, as estimated from the divergence of the
infinite-size localization (correlation) length $\xi(W)$ as a
function of the disorder strength $W$ at the transition $W=W_{\rm c}$, i.e.,
$\xi \propto |1-W/W_{\rm c}|^{-\nu}$, was systematically {\em
  underestimated}. The divergent nature at the transition could only
be poorly captured by FSS of data obtained for small system sizes and
large errors $\varepsilon$ in these finite-size data.  However, as
more powerful computers became available in the last decade, one
observed a trend towards larger values of $\nu \approx 1.35$
\cite{KraS96,SchKM89,KraBMS90,HofS93b} for $\varepsilon \leq 1\%$.

In 1994, high-precision data ($\varepsilon \leq 0.2\%$) of MacKinnon
\cite{Mac94} for the Anderson model of localization (AM) showed a
hitherto neglected systematic shift of the transition point $W_{\rm
  c}$ with increasing system size.  Taking this into account
phenomenologically, $\nu = 1.54\pm0.08$ was found \cite{Mac94}. A
subsequent approach by Slevin/Ohtsuki \cite{SleO99a,SleO99b,OhtSK99}
incorporated these shifts as a consequence of irrelevant scaling variables and further
allowed for corrections to scaling due to nonlinearities. With
higher-precision data ($\varepsilon\approx 0.1\%$), they found
$\nu=1.57\pm 0.04$.  Further results for, e.g., the AM with
anisotropic hopping \cite{MilRSU00,MilRS99a,MilRS01}, the off-diagonal
AM \cite{CaiRS99,BisCRS00}, and the AM in a magnetic field
\cite{ZhaK98,ZhaK97}, confirmed this value of $\nu$ within the error
bars \cite{CaiNRS01}. Also, $\nu$ is identical for the MIT as a
function of disorder or energy \cite{CaiRS99,BisCRS00}. We emphasize
that a properly performed Slevin/Ohtsuki scaling procedure needs
to assume various fit functions and that the final estimates are to be
suitably extracted from many such functional forms and starting
parameters \cite{MilRS99a,MilRS01,CaiRS99}; bootstrap
\cite{SleO99a,SleO99b,OhtSK99} or Monte Carlo methods
\cite{MilRS99a,MilRS01,CaiRS99} then need to be employed for a precise
estimate of error bars.

Regarding experiments, we note that similarly precise data ($0.1\%$)
are much harder to obtain for our experimental colleagues.
Nevertheless, recent advances in this direction based on careful
finite-temperature analysis of the conductivity data show a clear
trend towards increasing $\nu > 1$
\cite{StuHLM93,WafPL99,BogSB99,BogSB99b,ItoWOH99}. The roles of sample
inhomogeneities, magnetic effects and other possible experimental
influences are also discussed\cite{RosTP94,StuHLM94,Cas01}.

The statistical properties of spectra of disordered single-electron
systems are closely related to the localization properties of the
corresponding wave functions \cite{AbrALR79,AltS86,KraLAA94}.  In the
3D AM we have the insulating, the critical and the metallic phases,
respectively. For the insulating regime, the localized states even if
they are close in energy have an exponentially small overlap and their
levels are uncorrelated.  Accordingly, in the thermodynamic limit the
normalized distribution of the spacing $s$ between neighboring energy
levels follows the Poisson law
\begin{equation}
\label{eq-pois}
 P_{\rm P}(s)= \exp(-s).
\end{equation}
In the metallic regime, the large overlap of delocalized states
induces correlations in the spectrum leading to level repulsion. In
this case, if the system is invariant under rotational and under
time-reversal symmetry, the normalized spacing distribution closely
follows the Wigner surmise of the Gaussian orthogonal ensemble (GOE)
of random matrices \cite{Wig55,Wig57,Dys62,Meh90,Efe83},
\begin{equation}
\label{eq-wigner}
P_{\rm WD}(s)=\frac{\pi}{2}s \exp\left(-\frac{\pi}{4}s^2\right).
\end{equation}
The third symmetry class at the MIT is usually called the
critical statistics \cite{ShkSSL93,HofS94b,KraM97,ZhaK98}. Its
normalized level spacing distribution for large $s$ is proportional to
\begin{equation}
\label{eq-crit_chi}
P_{\rm c}(s)\propto \exp\left( -\kappa s \right)
\end{equation}
where $\kappa$ has been argued to be related to the value of the level
compressibility $\chi_{\rm c}$ at the MIT \cite{AltZSK88}.

Various measures have been suggested besides $P(s)$ as providing
alternative descriptions of the MIT depending on which theoretical and
numerical method is being used \cite{Met98,MilRS99a}. Of particular
interest is the so-called number-level variance $\Sigma_2$, which is a
measure of the {\em global} spectral rigidity \cite{Meh90}. It is
defined as
\begin{equation}
\label{eq-sigm}
\Sigma_{2}(\Delta E)=
\Bigl\langle\bigl[ n(\Delta E)-
\langle n(\Delta E)\rangle\bigr] ^{2}\Bigr\rangle
\end{equation}
where $n$ denotes the number of levels in a fixed energy interval
$\Delta E$ and $\langle\rangle$ indicates an averaging over disorder.
In the insulating state $\Sigma_2= \langle n \rangle$, while it has a
logarithmic increase $\Sigma_2\propto \ln \langle n \rangle$ in the
metallic state \cite{Meh90}. The behavior of the number variance at
the critical point has been conjectured to be Poisson-like
\cite{AltZSK88}, i.e., linear in $\langle n \rangle$,
\begin{equation}
\label{eq-si2}
\Sigma_2 \approx \chi \langle n \rangle
\end{equation}
where the level compressibility $\chi$ is another important parameter
to characterize the Anderson transition. It is defined as \cite{Met98,BogGS01}
\begin{equation}
\label{eq-si3}
\chi \approx \lim_{\langle n \rangle \rightarrow \infty} \
\lim_{N\rightarrow \infty} \frac{d\Sigma_2(\langle n \rangle)}{d
  \langle n \rangle}
\end{equation}
and takes values $0\le\chi\le1$, being zero in the metallic and unity in
the insulating state. It is a universal parameter and depends only on
the spatial dimensionality and on the symmetry class
\cite{ZhaK94,Mir00}. The two limits in Equation (\ref{eq-si3}) do not
commute. This non-commutativity is attributed to the fractal nature of
the critical states (see \cite{KraM97} and references therein).
The proposed linear increase (\ref{eq-si2}) of $\Sigma_2$ at the MIT
as a function of $\langle n \rangle$ at the transition has been a
matter of discussion \cite{ShkSSL93,KraLAA94}. In general, there is
consensus that $\Sigma_2$ has a {\em quasi-Poisson} behavior as in
Equation (\ref{eq-si2}) at the MIT
\cite{AroM95,KraLAA94,KraL95,AroKL94,BraMP98}.

In this paper we show how $\Sigma_2$ and $\chi$ can be used
together with FSS to obtain reliable estimates of the critical
exponent $\nu$. We employ various FSS schemes to check the
accuracy of our results. Our study goes beyond similar previous
investigations of the level number variance \cite{ZhaK94} due to a
considerably enhanced accuracy in the scaling data. Based on raw
spectral data of an anisotropic version of the AM, we find that
$\nu$ is consistently larger than $1$ in contradistinction to a
recently raised objection \cite{Sus01} to the FSS method of Ref.\
\cite{SleO99a}, cp.\ Appendix. The values of $\nu$ that we obtain
are in good agreement with the above mentioned recent estimates
for the isotropic case
\cite{Mac94,SleO97,MilRS01,CaiRS99,SleO99a,SleO99b,OhtSK99}. The
mean value of $\chi_{\rm c}$ at the MIT is $\approx 0.28 \pm
0.06$.

%%%%%%%%%%%%%%%%%%%%%%%%%%%%%%%%%%%%%%%%%%%%%%%%%%%%%%%%%%%%%%%%%%%%%%%%%%
\section{The Model Hamiltonian}
\label{sec-model}
%%%%%%%%%%%%%%%%%%%%%%%%%%%%%%%%%%%%%%%%%%%%%%%%%%%%%%%%%%%%%%%%%%%%%%%%%%

We consider the 3D Anderson model of localization described by a
Hamiltonian in the lattice site basis as
\begin{equation}
\label{eq-ham}
H=\sum_{i}\varepsilon_{i}|i \rangle \langle i|+\sum_{i \neq j}t_{ij}|i
\rangle \langle j|.
\end{equation}
The states $|i{\rangle}$ are orthonormal and correspond to particles
located at the $N^3$ sites $i=(x,y,z)$ of a regular cubic lattice with
periodic boundary conditions. The site energies $\varepsilon_i$ are
taken to be random numbers uniformly distributed in the interval
$[-W/2,W/2]$; $W$ defines the disorder strength.
The hopping integrals $t_{ij}$ are restricted to nearest neighbors and
depend only on the three spatial directions.  In this paper we
consider weakly coupled planes defined by $t_{x}=t_{y}=1$,
$t_{z}=0.1$.
We emphasize that we have chosen the strong anisotropy simply because
we have the most accurate data (the relative error ranges from $0.2$
to $0.4\%$) available for this value from a previous study
\cite{ZamLES96a,MilRS99a,Mil00}. This high accuracy (for spectral
data) has been achieved by averaging over $10$ samples for system size
$50^3$ and then increasing the number of samples up to $699$ for
system size $13^3$ such that always at least $10^5$ eigenenergies have
been computed for each $N$ and $W$. Since it was shown in Refs.\
\cite{MilRS99a,Mil00} numerically that the universality class of the
model is not changed by the anisotropy, we therefore need not generate
similarly precise data for the isotropic model in order to show
scaling of $\Sigma_2$ and $\chi$.

The Hamiltonian (\ref{eq-ham}) was diagonalized numerically using a
Lanczos method \cite{CulW85a}. In order to perform any
statistical calculations the eigenspectrum is "unfolded" so that the
average spacing between adjacent eigenvalues is one. Spectra unfolding
amounts to a kind of renormalization of the eigenvalues in order to
extract the universal spectral properties. One way to perform spectral
unfolding is to subtract the regular part from the integrated density
of states and consider only the fluctuations \cite{Meh90}. This can be
achieved by different means; however, there is no rigorous
prescription and the ``best'' criterion is the insensitivity of the
final result to the method employed. This criterion is fulfilled in
the present study.

%%%%%%%%%%%%%%%%%%%%%%%%%%%%%%%%%%%%%%%%%%%%%%%%%%%%%%%%%%%%%%%%%%%%%%%%%%
\section{Finite-Size Scaling}
\label{sec-fss}
%%%%%%%%%%%%%%%%%%%%%%%%%%%%%%%%%%%%%%%%%%%%%%%%%%%%%%%%%%%%%%%%%%%%%%%%%%

According to the one-parameter-scaling hypothesis \cite{AbrALR79}, a
quantity $X$ at different disorders $W$ and energies $E$ scales onto a
single scaling curve, i.e.,
\begin{equation}
  \label{eq-scaling}
  X(N;W,E) = f(\xi(W,E)/N),
\end{equation}
where the scaling function $f$ is a generalized homogeneous function
\cite{Sta87} and $\xi$ denotes the correlation length. The MIT in the
3D AM is a second-order phase transition and as such it is
characterized by a divergent correlation length with power-law
behavior $\xi_{\infty}=|W-W_{\rm c}|^{-\nu}$. The task of FSS now is
to determine the infinite-size quantities $f$ and $\xi(W,E)$ from
finite-size data and to obtain the critical exponent $\nu$ and the
critical disorder $W_{\rm c}$ or the critical energy $E_{\rm c}$.

The essential idea of the FSS procedure of Ref.\ \cite{SleO99a} is to
construct a family of fit functions which include corrections to
scaling due to an irrelevant scaling variable and due to
non-linearities of the disorder dependence of the scaling variables.
The former is only necessary when the accuracy of the data allows us
to observe systematic shifts of the intersection points for different
$W$ (or $E$) and $N$. In all current FSS studies of spectral
properties such an accuracy has not been reported, only studies using
the transfer-matrix method allow for an identification of irrelevant
variables.

Following Refs.\ \cite{SleO99a,MilRS99a,MilRS01}, we thus assume a
scaling form without irrelevant variables to be
\begin{equation}
  \label{eq-Slevin}
%  X=\tilde{f}(\chi_{\rm r} N^{1/\nu}, \chi_{\rm i} N^{y})
  X=\tilde{f}(g_{\rm r} N^{1/\nu}),
\end{equation}
where $g_{\rm r}$ is the relevant scaling variable. Taylor expanding
$\tilde{f}$ up to order $n_{\rm r}$, we get
\begin{equation}
  \label{eq-Slevin-ftilde}
  \tilde{f}=\sum_{i=0}^{n_{\rm r}} b_{i} g_{\rm r}^i N^{i/\nu} .
\end{equation}
Non-linearities are taken into account by expanding $g_{\rm r}$ in
terms of $u=1-W/W_{\rm c}$ (or $u=1-E/E_{\rm c}$) up to order $m_{\rm
  r}$
\begin{equation}
  \label{eq-Slevin-Var}
  g_{\rm r}(u)=u + \sum_{k=2}^{m_{\rm r}} b_k u^k.
\end{equation}
The fit function is adjusted to the data by choosing the orders
$n_{\rm r}$ and $m_{\rm r}$ up to which the expansions are carried
out.

%%%%%%%%%%%%%%%%%%%%%%%%%%%%%%%%%%%%%%%%%%%%%%%%%%%%%%%%%%%%%%%%%%%%%%%%%%
\section{Results}
\label{sec-results}
%%%%%%%%%%%%%%%%%%%%%%%%%%%%%%%%%%%%%%%%%%%%%%%%%%%%%%%%%%%%%%%%%%%%%%%%%%

%%%%%%%%%%%%%%%%%%%%%%%%%%%%%%%%%%%%%%%%%%%%%%%%%%%%%%%%%%%%%%%%%%%%%%%%%%
\subsection{The spectral rigidity}
\label{sec-rigidity}

\begin{figure}
  \resizebox{0.49\textwidth}{!}{ \includegraphics{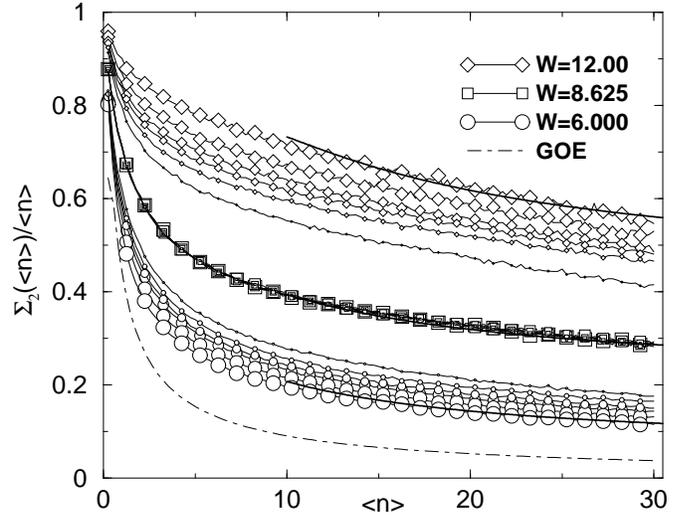}}
\caption{\label{fig-a}
  $\Sigma_2(\langle n \rangle)/\langle n \rangle$ for coupled planes
  with $t_z=0.1$ and for linear system sizes $N=13, 17, 21, 24, 30$ and
  $40$ denoted by increasing symbol size. The dashed line correspond to
  the GOE. Note that only every $5$th symbol is shown. The
  three thick solid lines indicate fits according to Equation
  (\ref{eq-invpolyfit}) of data for $ \langle n \rangle > 15$.}
\end{figure}

In Figure \ref{fig-a} we show the computed $\Sigma_2$ data for, e.g.,
three disorders, $E\in [-4.1,4.1]$ ($50\%$ of the spectrum), and
various system sizes. In a previous study \cite{MilRS99a}, we have
shown that similar level-statistics results can be obtained when only
$20\%$ of the spectrum close to the band center are taken into account
\cite{SirP98}.  The dependence of $W_{\rm c}$ on energy has been
considered previously in, e.g.\ Ref.\ \cite{KraBMS90}. A large band of
states around $E=0$ shows the same multifractal characteristics as a
narrow band \cite{MilRS97}. Thus it is justified to take a large part
of the spectrum into account when computing spectral statistics.
It is evident from the figure that there is a systematic size
dependence as a function of disorder. For large disorder $W=12$ and
upon increasing the system size the data approach the insulating
(Poisson) behavior.  Similarly, for small disorder $W=6$ the curves
tend towards the metallic (GOE) behavior. And close to the MIT at
$W_{\rm c}\approx 8.625$, the data for all system sizes collapse onto
a single curve. A similar trend as in Figure \ref{fig-a} has been
observed for $\Sigma_2$ in the four-dimensional isotropic AM
\cite{ZhaK98}.
%%%%%%%%%%%%%%%%%%%%%%%%%%%%%%% FIGURE %%%%%%%%%%%%%%%%%%%%%%%%%%%%%%%%%%%%%%
%%%%%%%%%%%%%%%%%%%%%%%%%%%%%%% END OF FIGURE %%%%%%%%%%%%%%%%%%%%%%%%%%%%%%%

%%%%%%%%%%%%%%%%%%%%%%%%%%%%%%%%%%%%%%%%%%%%%%%%%%%%%%%%%%%%%%%%%%%%%%%%%%
\subsection{FSS with integrated $\Sigma_2$ data}
\label{sec-eta}

\begin{figure}
\resizebox{0.49\textwidth}{!}{
\includegraphics{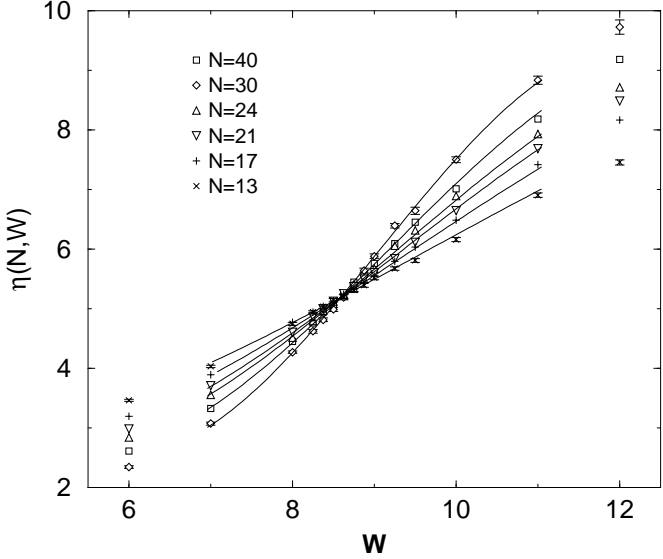}}
\caption{\label{fig-b}
  Integrated $\Sigma_2$, $\eta(N,W)$, for coupled planes with
  $t_z=0.1$ for linear system sizes shown. The solid lines are fit
  functions of FSS from Equation (\ref{eq-Slevin}) with $n_r=3$ and
  $m_r=1$. }
\end{figure}

\begin{figure}
\resizebox{0.49\textwidth}{!}{
\includegraphics{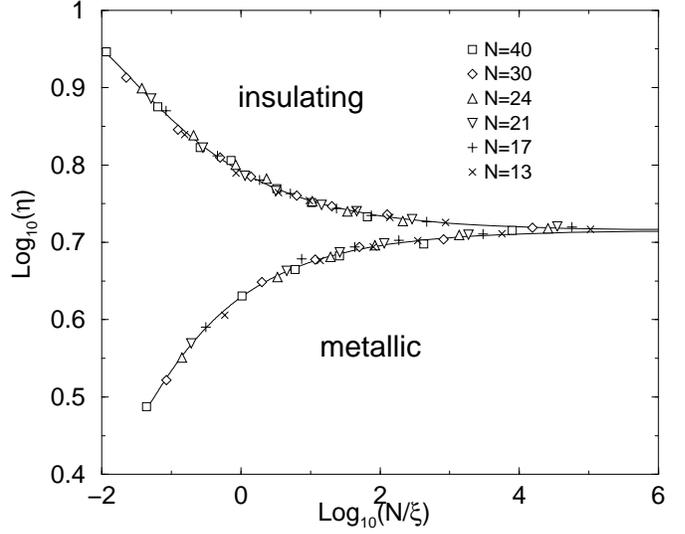}}
\caption{\label{fig-bb}
  The one-parameter scaling dependence of $\eta$ on $\xi$ for
  different system sizes $N$ and disorders $W \in [6,12]$.}
\end{figure}

In order to perform FSS, we could now use the data of Figure
\ref{fig-a} and plot them at each value of $\langle n \rangle$ as a
function of disorder. However, such an approach is of limited
usefulness since it is apriori unclear how to weigh data from
different $\langle n \rangle$ values. Furthermore, the fluctuations in
the data lead to rather large error bars in the obtained estimates of
$\nu$ and $W_{\rm c}$.
Instead, we define the integrated quantity
\begin{equation}
  \label{eq-eta}
  \eta(N,W)=\frac{1}{L_{0}}\int_{0}^{L_{0}}\Sigma_{2}(L)dL
\end{equation}
with $L=\langle n \rangle$. This is similar to the FSS analysis of
$\Delta_{3}$-statistics \cite{Mil00}. The integral is also considered
up to $L_{0}=30$ only because the relative error in $\Sigma_2$ becomes
rather large for larger $L_{0}$ values and hence the calculation is
less reliable. It is evident from Figure \ref{fig-b} that $\eta(N,W)$
shows the desired system-size dependence for various values of $W$
exhibiting insulating, critical and metallic behavior for $W$ larger,
close to, and smaller than $W_{\rm c}\approx 8.625$.
%%%%%%%%%%%%%%%%%%%%%%%%%%%%%%% FIGURE %%%%%%%%%%%%%%%%%%%%%%%%%%%%%%%%%%%%%%
%%%%%%%%%%%%%%%%%%%%%%%%%%%%%%% END OF FIGURE %%%%%%%%%%%%%%%%%%%%%%%%%%%%%%%
In order to obtain $\xi$ from finite system-size data we now use the FSS
procedure of Section \ref{sec-fss}.  For the non-linear fit, we used the
Levenberg-Marquardt method \cite{SleO99a}. In Figure \ref{fig-bb} we
show that the data from different system sizes collapse on two branches
corresponding to localized and extended behavior.  This clearly shows
that $\eta$ exhibits one-parameter FSS. We then compute the critical
exponent $\nu$ and the critical disorder $W_{\rm c}$ for various
parameters. The results are tabulated in Table\ \ref{tab-eta}.  The
average values are $\nu \approx 1.43 \pm 0.13$ and $W_{\rm c} \approx
8.62 \pm 0.04$, respectively. Here and in the following, the error
intervals are standard errors, i.e., denoting {\em one} standard
deviation.
%%%%%%%%%%%%%%%%%%%%%%%%%%%%%%% FIGURE %%%%%%%%%%%%%%%%%%%%%%%%%%%%%%%%%%%%%%
%%%%%%%%%%%%%%%%%%%%%%%%%%%%%%% END OF FIGURE %%%%%%%%%%%%%%%%%%%%%%%%%%%%%%%
%%%%%%%%%%%%%%%%%%%%%%%%%%%%%%% TABLE %%%%%%%%%%%%%%%%%%%%%%%%%%%%%%%%%%%%%%%
\begin{table}[tbh]
\caption{\label{tab-eta}
  Parameters for FSS of $\eta$ and the resulting estimates for $\nu$ and
  $W_{\rm c}$. The numbers in the $3$rd and $4$th column denote orders
  $n_{\rm r}$ and $m_{\rm r}$ used in the expansions (10) and (11),
  respectively, for which the best fits have been obtained. Q is the
  goodness-of-fit defined as usual by the $\chi^2$ quality of fit
  \cite{PreTVF92} and quoted errors correspond to one standard
  deviation.}
\begin{tabular}[h]{|c|c|c|c|c|c|c|}
\hline\noalign{\smallskip}
   $W$ & $N$ & $n_{\rm r}$ & $m_{\rm r}$ & $Q$ & $W_{\rm c}$ & $\nu$\\
\hline
   $7\cdots11.0$ & $13\cdots40$ & $3$ & $1$  &$0.01$& $8.59 (3)$ & $1.38 (3)$ \\
   $7\cdots11.0$ & $13\cdots40$ & $1$ & $3$  &$0.09$& $8.56 (4)$ & $1.57 (4)$ \\
   $8\cdots9.25$ & $13\cdots40$ & $3$ & $1$  &$0.70$& $8.62 (3)$ & $1.35 (12)$\\
   $8\cdots9.25$ & $13\cdots40$ & $1$ & $3$  &$0.54$& $8.61 (3)$ & $1.33 (9)$ \\
   $8\cdots9.25$ & $13\cdots40$ & $3$ & $2$  &$0.75$& $8.64 (4)$ & $1.35 (13)$\\
   $8\cdots9.25$ & $13\cdots40$ & $2$ & $3$  &$0.79$& $8.64 (2)$ & $1.27 (10)$\\
   $8\cdots9.25$ & $24\cdots40$ & $3$ & $3$  &$0.76$& $8.64 (3)$ & $1.29 (15)$\\
   $7\cdots11.0$ & $24\cdots40$ & $3$ & $1$  &$0.14$& $8.63 (6)$ & $1.48 (13)$\\
   $7\cdots11.0$ & $24\cdots40$ & $1$ & $3$  &$0.02$& $8.61 (8)$ & $1.80 (18)$\\
   $8\cdots9.25$ & $24\cdots40$ & $3$ & $1$  &$0.72$& $8.66 (6)$ & $1.50 (26)$\\
   $8\cdots9.25$ & $24\cdots40$ & $1$ & $3$  &$0.67$& $8.64 (5)$ & $1.47 (22)$\\
   $7\cdots11.0$ & $13\cdots40$ & $3$ & $3$  &$0.03$& $8.60 (3)$ & $1.34 (7) $\\[1ex]\hline\multicolumn{5}{|l|}{average:}
&$8.62 (4)$ & $1.43 (13)$\\
\noalign{\smallskip}\hline
\end{tabular}
\end{table}
%%%%%%%%%%%%%%%%%%%%%%%%%%%%%%% END OF TABLE %%%%%%%%%%%%%%%%%%%%%%%%%%%%%%%%

%%%%%%%%%%%%%%%%%%%%%%%%%%%%%%%%%%%%%%%%%%%%%%%%%%%%%%%%%%%%%%%%%%%%%%%%%%
\subsection{FSS with $\chi$}
\label{sec-chi}

We now turn our attention to computing $\chi$. First we note that
$\Sigma_{2}(L)/L$ ($L\equiv\langle n \rangle$) as plotted in Figure
\ref{fig-a} is already a crude approximation of $\chi$. Since there is
a systematic size dependence, this already indicates that $\chi$
should obey FSS. In order to proceed more accurately, we now fit the
${\Sigma_2(L)}/{L}$ data with an ansatz function containing irrelevant
scaling exponents $y_k$, i.e.\
\begin{equation}
\label{eq-invpolyfit}
{\frac{\Sigma_{2}(N,W,L)}{L}}\approx
\chi + \sum_{k=1}^{m} a_{k}(N,W)L^{-y_k}
\end{equation}
up to order $m$. Thus in the limit $L \rightarrow \infty$, the
constant term will be equal to the desired value of the level
compressibility $\chi$. The data used in the fits range from $L=0$ up
to $L\approx 140$. Data for larger $L$ values ($\approx 250$) was
ignored due to reduced statistical accuracy. In Figure \ref{fig-a}, we
show some typical fits for large system sizes.

We next perform FSS of $\chi$ as explained above. However, a
non-linear-fit procedure of (increasingly fluctuating) large $L$-data
with exponents as fitting parameters is inherently unstable. Thus it
is numerically much better to fit with fixed exponents.  Using such a
fit with $W= 7, \ldots, 11$, $N= 24, \ldots, 50$ for $m=1$, we find
that $\chi$ ranges from $0.275$ to $0.292$ with $y_1$ varying from
$0.95$ to $1.3$, respectively.  Various other combinations for values
of $y_1= 0.95, \ldots, 1.3$ and $y_2= 1.5,\ldots, 2.5$ result in the
averaged FSS estimate $\chi_{\rm c}= 0.28 \pm 0.03$. Other values
outside these ranges do not fit ${{\Sigma_{2}(N,W,L)}/{L}}$ and
larger values for $m$ do not enhance the quality of the FSS.

\begin{figure}
\resizebox{0.49\textwidth}{!}{
\includegraphics{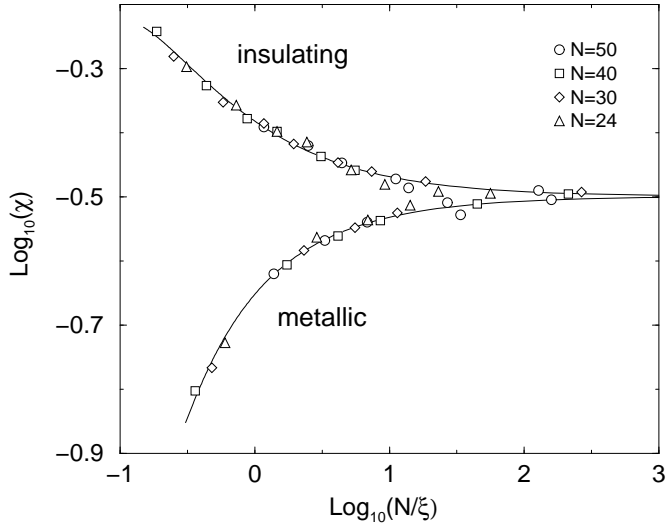}}
\caption{\label{fig-LONG}
  The one-parameter scaling dependence of $\chi$, obtained from
  Equation (\protect\ref{eq-invpolyfit}) for, e.g., $y_1=1$ and
  $y_2=2$, on $\xi$ for disorders $W \in [7,11]$, system sizes $24$,
  $30$, $40$, and $50$, with $n_{\rm r}=3$, $m_{\rm r}= 1$. The dashed
  line indicates the value $\chi_{\rm c}=0.27(1)$ at the MIT obtained
  from this fit.}
\end{figure}

As shown, e.g.\ for $y_1=1$ and $y_2=2$ in Figure \ref{fig-LONG},
$\chi$ data for different system sizes and $W$ collapse onto a single
scaling curve with two branches.  From this and further data for
different ranges of $W$ and $N$, we can roughly estimate $\chi_{\rm
  c}$ at the MIT to be $0.28 \pm 0.06$ as shown in Table
\ref{tab-chi}. This value is in good agreement with previously
obtained estimates \cite{ZhaK95b,Met98,Can96}. We also note that
Equation (\ref{eq-crit_chi}) with $\kappa= 1/2\chi_{\rm c}\approx 1.8$
fits the large-$s$ tails of $P(s)$ at the MIT reasonably well.
%%%%%%%%%%%%%%%%%%%%%%%%%%%%%%% FIGURE %%%%%%%%%%%%%%%%%%%%%%%%%%%%%%%%%%%%%%
%%%%%%%%%%%%%%%%%%%%%%%%%%%%%%% END OF FIGURE %%%%%%%%%%%%%%%%%%%%%%%%%%%%%%%

Unfortunately, the fit in $L^{-y_k}$ is not good enough to reproduce
the values for $\nu$ and $W_{\rm c}$ with the desired high accuracy as
shown, e.g., in Table \ref{tab-chi}. This is because the data for
large $L$ fluctuate much more strongly than at small $L$ due to the
reduced statistics of such extremely large spacings. E.g., even for a
system of size $50^3$, we have only 250 spacings with $L=250$
available when taking only the central half of the spectrum into
consideration to avoid distortions from the localized states in the
band tails.  Furthermore, the usual unfolding procedures rely on local
spectral interpolations and may no longer work for such large
spacings. In fact, using an unfolding suitable for small $L$
statistics, we could erroneously reduce the estimated value of
$\chi_{\rm c}$ to $\approx 0.25$.

%%%%%%%%%%%%%%%%%%%%%%%%%%%%%%%%%%%%%%%%%%%%%%%%%%%%%%%%%%%%%%%%%%%%%%%%%%
\subsection{FSS with $\chi$ from truncated data.}
\label{sec-trachi}

\begin{figure}
\resizebox{0.49\textwidth}{!}{
\includegraphics{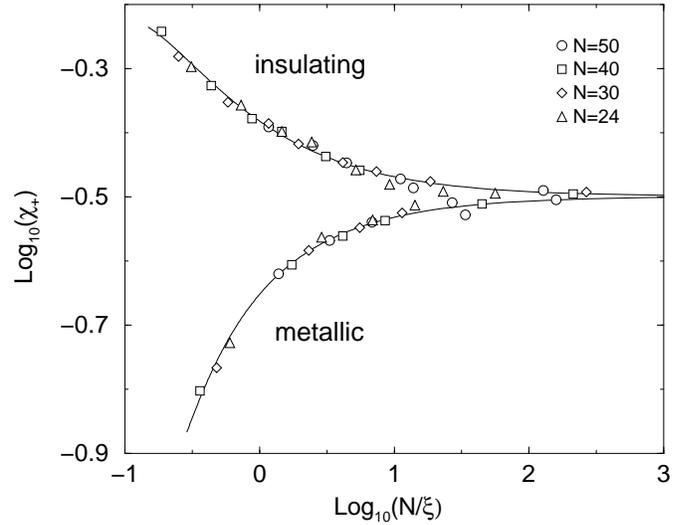}}
\caption{\label{fig-SHORT}
  The one-parameter scaling dependence of $\chi_+$, obtained from
  Equation (\protect\ref{eq-invpolyfit}) for $y_1=1$ and $y_2=2$
  fitted up to $L=30$, on $\xi$ for disorders $W \in [7,11]$, system
  sizes $24$, $30$, $40$, and $50$, with $n_{\rm r}=3$, $m_{\rm r}=
  1$.}
\end{figure}

In order to suppress the problems with large-$L$ fluctuations, we have
truncated the $\Sigma_2/L$ data at $L=30$ and performed the FSS
procedure as before with Equation (\ref{eq-invpolyfit}) using fixed
$y_1=1$ and $y_2=2$.  As shown in Figure \ref{fig-SHORT}, the $\chi$
data for different system sizes and different $W$ values collapse
again onto a single scaling curve with two branches.
Due to the truncation, the estimated values $\chi_{+}$ can only serve
as upper limits to the true value $\chi_{\rm c}$ at the MIT. But the
resulting values for $\nu$ and $W_{\rm c}$ are of much better accuracy
and are shown in Table \ref{tab-chi}.  For the average critical
exponent we obtain $\nu \approx 1.44 \pm 0.13$ and for the average
critical disorder $W_{\rm c} \approx 8.66 \pm 0.04$.

%%%%%%%%%%%%%%%%%%%%%%%%%%%%%%% FIGURE %%%%%%%%%%%%%%%%%%%%%%%%%%%%%%%%%%%%%%
%%%%%%%%%%%%%%%%%%%%%%%%%%%%%%% END OF FIGURE %%%%%%%%%%%%%%%%%%%%%%%%%%%%%%%

%%%%%%%%%%%%%%%%%%%%%%%%%%%%%%%%%%%%%%%%%%%%%%%%%%%%%%%%%%%%%%%%%%%%%%%%%%
\subsection{FSS with a polynomial fit}
\label{sec-polchi}

\begin{figure}
\resizebox{0.49\textwidth}{!}{
\includegraphics{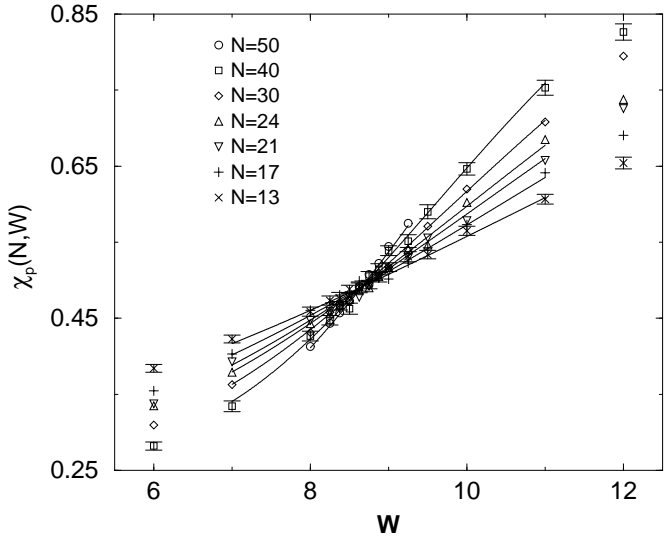}}
\caption{\label{fig-d}
  $\chi_{\rm p}(N,W)$ obtained by fitting the $\Sigma_{2}$ data with
  $m=3$ in Equation (\ref{eq-polyfit}).  System-size dependence is
  clearly seen.  The solid lines are fit functions from FSS of
  Equation (\ref{eq-Slevin}) with $n_r=3$ and $m_r=1$.}
\end{figure}

In order to proceed more accurately with the determination of $\nu$
and $W_{\rm c}$, we now fit $\Sigma_2(L)$ data for small $L$ with a
polynomial in $L$, i.e.,
\begin{equation}
\label{eq-polyfit}
\frac{\Sigma_{2}(N,W,L)}{L}\approx
\chi_{\rm p} + \sum_{k=1}^{m} b_{k}(N,W)L^{k}
\end{equation}
up to order $m$. We then identify a rough estimate of the level
compressibility with the linear expansion coefficient $\chi_{\rm p}$.
This implies a systematic shift of $\chi$ and the value of $\chi_{\rm
  p}$ at the MIT will also be shifted towards a larger value when
compared to $\chi_{\rm c}$.  However, $\nu$ and $W_{\rm c}$ can be
determined with increased precision: the quality of the fit, cp.\
Figure \ref{fig-d}, is very good and certainly better than in the two
previous cases.  As a check to the numerical reliability of this
method, we vary the value of $L$ included in the fit function
(\ref{eq-polyfit}) by fitting the $\Sigma_2/L$ data for various ranges
of $L =10,15,20,25 $ and $30$. We find that there is only a negligible
change in the obtained values of $\nu$ and $W_{\rm c}$.
%%%%%%%%%%%%%%%%%%%%%%%%%%%%%%% FIGURE %%%%%%%%%%%%%%%%%%%%%%%%%%%%%%%%%%%%%%
%%%%%%%%%%%%%%%%%%%%%%%%%%%%%%% END OF FIGURE %%%%%%%%%%%%%%%%%%%%%%%%%%%%%%%

\begin{figure}
\resizebox{0.49\textwidth}{!}{
\includegraphics{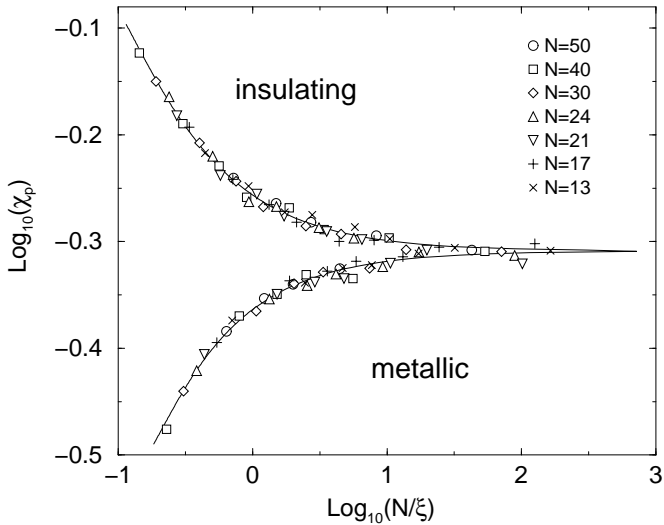}}
\caption{\label{fig-db}
  The one-parameter scaling dependence of $\chi_{\rm p}$ on $\xi$ for
  different system sizes $N$ and disorders $W \in [6,12]$.}
\end{figure}

After FSS, $\chi_{\rm p}$ data for {\em all} system sizes and {\em
  all} $W$ collapse onto two curves as shown in Figure \ref{fig-db}.
Results for $\nu$ and $W_{\rm c}$ for various FSS functions and
different $m$ are shown in Table\ \ref{tab-chibar}.  For the average
critical exponent we obtain $\nu \approx 1.47 \pm 0.10$ and for the
average critical disorder $W_{\rm c} \approx 8.56 \pm 0.05$.
%%%%%%%%%%%%%%%%%%%%%%%%%%%%%%% FIGURE %%%%%%%%%%%%%%%%%%%%%%%%%%%%%%%%%%%%%%
%%%%%%%%%%%%%%%%%%%%%%%%%%%%%%% END OF FIGURE %%%%%%%%%%%%%%%%%%%%%%%%%%%%%%%

The values of $\nu$ calculated from the $\eta$- and the three
$\chi$-based approaches are compatible with each other and are also
comparable to values from other methods \cite{Sch99a,SleMO01}, for
instance the transfer-matrix method which gives $\nu \approx 1.62 \pm
0.07$ \cite{MilRSU00}. We can therefore claim that both $\chi$ and
$\eta$ are good FSS parameters to characterize the MIT. Nevertheless, a
simple fitting procedure in the large $L$ limit, although in principle
correct, will encounter many numerical problems.

%%%%%%%%%%%%%%%%%%%%%%%%%%%%%%%%%%%%%%%%%%%%%%%%%%%%%%%%%%%%%%%%%%%%%%%%%%
\section{Conclusion}
\label{sec-conclusion}
%%%%%%%%%%%%%%%%%%%%%%%%%%%%%%%%%%%%%%%%%%%%%%%%%%%%%%%%%%%%%%%%%%%%%%%%%%

States at the 3D MIT are multifractal entities \cite{SchG91,WEH2000}.
This implies that, while not being extended, their spatial structure
nevertheless results in a long-ranged, power-law overlap of electronic
densities in energy \cite{ChaKL96,KraM97}, i.e.,
\begin{equation}
  \label{eq-density-corr}
  \langle |\Psi_i|^2 |\Psi_k|^2 \rangle \propto |E_i - E_k|^{-(1-D_2/d)}
\end{equation}
where $D_2$ is the correlation dimension \cite{MilRS97} and the
connection $\chi_{\rm c}= (d-D_{2})/2d$ has been conjectured
\cite{ChaKL96}. In order to describe generic features of such
multifractal states, various {\em critical} random matrix models have
been suggested and studied
\cite{MosNS94,KraM97,AltL97,Mir96,MutCIN93}, albeit mostly for the
unitary class of models.
Using the above relation of $D_{2}$ (see Refs.\ \cite{MilRS97,Mil00}
for numerical estimates at the MIT in anisotropic 3D
AMs) with $\chi_{\rm c}$, we find that the computed value $\chi_{\rm
  c}= 0.28 \pm 0.06$ is compatible with $D_{2}\approx 1.3\pm 0.1$, which can be
calculated easily from the $f(\alpha)$ spectra published in Refs.\
\cite{MilRS97,Mil00,SchMRE99}.
On the other hand, it has been shown that in the limit of ``strong
multifractality'', the above connection between $\chi_{c}$ and $D_{2}$
no longer holds \cite{EveM00}.  Previous estimates of $D(2)$ in the
isotropic 3D AM range from $1.4$--$1.7$
\cite{BraHS96,BraMP98,ChaKL96,KawKO99,OhtK96,ZhoZSP00}.  Certainly,
our multifractal \cite{Mil00,MilRS97} is not an infinitely sparse
multifractal wave ($D(2)=0)$) as sometimes expected for the critical
ensembles \cite{KraM97}.

In summary, our results show that $\chi$ (and $\eta$) can indeed be used
to compute, with the help of FSS, estimates of $\chi_{\rm c}$, $W_{\rm
c}$ and $\nu$ which are in good agreement with transfer-matrix and other
spectral analysis. We are confident that the remaining small difference
in values can be further shrunk when larger system sizes become
available for the spectral statistics.

\section*{Acknowledgment}
We thank F.\ Milde for some of the data used in our calculations.  We
also thank F.\ Evers and V.\ E.\ Kravtsov for stimulating discussions.
Financial support from the Deutsche Forschungsgemeinschaft via SFB393 is
gratefully acknowledged.

\appendix
\section{Another FSS procedure}
\label{sec-another}

In a recent communication to the cond-mat archives \cite{Sus01}, the
FSS method used in the present paper has been criticized and the
results obtained by various groups
\cite{SleO99a,SleO99b,OhtSK99,MilRSU00,MilRS99a,MilRS01,CaiRS99,BisCRS00,ZhaK98,ZhaK97}
for the critical exponent $\nu$ of the localization length at the MIT
in the 3D AM have been questioned.  These claims are based on the
observation that there still is some disagreement between analytical,
numerical and experimental results for the critical exponent
\cite{KraM93}. Ref.\ \cite{Sus01} proposes yet another procedure to
deal with corrections to scaling. Furthermore, it is hinted that the
numerical data support $\nu\approx 1$, whereas the present manuscript
and recent numerical papers find $\nu \approx 1.5\pm 0.2$
\cite{SleO99a,SleO99b,OhtSK99}.

\begin{figure}[t]
\resizebox{0.49\textwidth}{!}{
\includegraphics{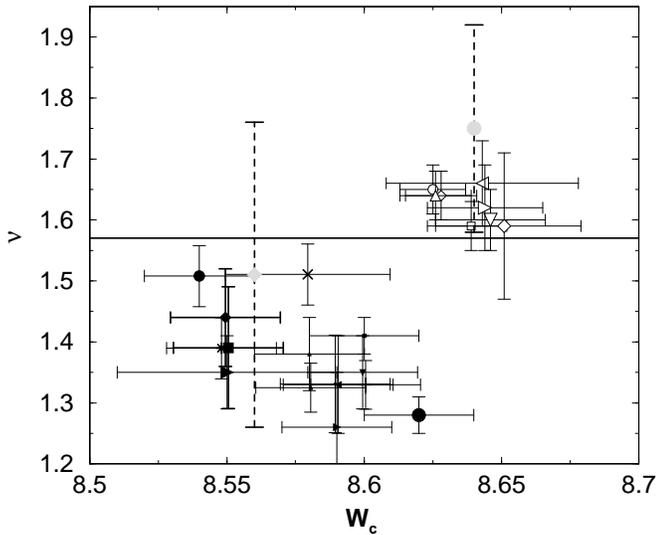}}
\caption{\label{fig-nu-wc}
  Results for $W_c$ and $\nu$, for the {\em anisotropic} AM
  \protect\cite{MilRSU00,MilRS99a} using FSS of Ref.\ \cite{SleO99a} for
  transfer-matrix data (open symbols) and level-statistics data (filled
  symbols) for various fit functions. The error bars show the $95\%$
  confidence intervals. The accuracy of the localization length data
  obtained from the transfer matrix method is an order of magnitude
  higher than that of the energy-level-statistics data. The system sizes
  of transfer-matrix data are larger than for level-statistics data,
  giving systematically larger $\nu$ values for the former. The goodness
  of a fit is reflected in the size of the symbol. The $2$ thick error
  bars mark high quality fits of level-statistics data for large system
  sizes.
  The gray $\circ$ and $\Box$ and the corresponding dashed error bars
  represent the result of the fit procedure of Ref.\ \cite{Sus01} applied
  to transfer-matrix data and level-statistics data for the anisotropic
  AM, respectively. The solid line marks the result of Ref.\ \cite{SleO99a}.}
\end{figure}

We have tested the method proposed by Ref.\ \cite{Sus01} first with
transfer-matrix data \cite{MilRSU00,CaiRS99,BisCRS00} with
$\varepsilon\leq 0.1\%$; we find
%%%%$\nu_{\mbox{\footnotesize\protect\cite{Sus01}}}= 1.75\pm 0.17$ for the
$\nu = 1.75\pm 0.17$ for the anisotropic and $1.55\pm 0.04$ for the
random-hopping AM.  The FSS of section \ref{sec-fss} gives $\nu=1.61
\pm 0.07$ \cite{MilRSU00} and $\nu=1.54\pm0.03$
\cite{CaiRS99,BisCRS00}, respectively, for the same set of data. Note
that the first value ($1.75$) is so high because systematic shifts of
$W_{\rm c}$ due to an irrelevant scaling variable are not taken into
account in \cite{Sus01}.
Using for a second test the energy-level-statistics data of the
present manuscript with $\varepsilon\approx 1\%$, we find
$\nu=1.51 \pm 0.25$.
Last, for artificially generated data with precisely known $W_c=16.5$
and varying $\nu\in [0.5,2.0]$ the results of the method of Ref.\
\cite{Sus01} are comparable to the results of the Kramer/MacKinnon FSS
\cite{MacK83} and slightly less reliable than the present FSS as shown in
Figure \ref{fig-nu-wc}.
%%%%%%%%%%%%%%%%%%%%%%%%%%%%%%% FIGURE %%%%%%%%%%%%%%%%%%%%%%%%%%%%%%%%%%%%%%
%%%%%%%%%%%%%%%%%%%%%%%%%%%%%%% END OF FIGURE %%%%%%%%%%%%%%%%%%%%%%%%%%%%%%%
We conclude that the method proposed in Ref.\ \cite{Sus01} also yields
$\nu \approx 1.58$ and not $\nu\approx 1$ for the MIT of the AM.

%%%%%%%%%%%%%%%%%%%%%%%%%%%%%%%%%%%%%%%%%%%%%%%%%%%%%%%%%%%%%%%%%%%%%
%\bibliographystyle{prsty}
%\bibliography{bibliograph}

%%%%%%%%%%%%%%%%%%%%%%%%%%%%%%% TABLE %%%%%%%%%%%%%%%%%%%%%%%%%%%%%%%%%%%%%%%
%\widetext
\newpage
\begin{table}[tbh]
\caption{\label{tab-chi}
  Parameters for FSS of $\chi$ as in (\ref{eq-invpolyfit}) with
  $y_1=1$, $y_2=2$ and the resulting estimates for $\chi_{\rm c}$,
  $\nu$ and $W_{\rm c}$. The numbers in the $3$rd and $4$th column
  denote orders $n_{\rm r}$ and $m_{\rm r}$ used in the expansions
  (\ref{eq-Slevin-ftilde}) and (\ref{eq-Slevin-Var}), respectively,
  for which the best fits have been obtained. The $5$th column is for
  fits based on $\Sigma_2$ data up to $L=140$, the $6-8$th columns are
  for fits up to $L=30$ only. The goodness-of-fit $Q$ \cite{PreTVF92}
  is less than $0.1$ in all cases.}
\begin{tabular}[h]{|c|c|c|c||c||c|c|c|}
\hline\noalign{\smallskip}
 & & & & $L_{\rm max}=140$ & \multicolumn{3}{c|}{$L_{\rm max}=30$} \\
   $W$ & $N$ & $n_{\rm r}$ & $m_{\rm r}$ & $\chi$ & $\chi_{+}$ & $W_{\rm c}$ & $\nu$\\
\hline
   $7\cdots11.0$ & $13\cdots50$ & $3$ & $1$  & $0.27 (9) $  & $0.31 (5)$& $8.69 (3)$ & $1.44 (7) $\\
   $7\cdots11.0$ & $13\cdots50$ & $1$ & $3$  & $0.16 (3) $  & $0.31 (3)$& $8.69 (3)$ & $1.58 (8) $\\
   $8\cdots9.25$ & $13\cdots50$ & $3$ & $1$  & $0.26 (12)$  & $0.30 (5)$& $8.69 (4)$ & $1.38 (16)$\\
   $8\cdots9.25$ & $13\cdots50$ & $1$ & $3$  &        $--$  & $0.31 (5)$& $8.68 (3)$ & $1.44 (14)$\\
   $8\cdots9.25$ & $13\cdots50$ & $3$ & $2$  & $0.31 (1) $  & $0.33 (1)$& $8.70 (4)$ & $1.35 (15)$\\
   $8\cdots9.25$ & $13\cdots50$ & $2$ & $3$  & $0.37 (2) $  & $0.37 (6)$& $8.70 (3)$ & $1.32 (16)$\\
   $8\cdots9.25$ & $24\cdots50$ & $3$ & $3$  & $0.27 (1) $  & $0.32 (1)$& $8.60 (3)$ & $1.11 (16)$\\
   $7\cdots11.0$ & $24\cdots50$ & $3$ & $1$  & $0.27 (1) $  & $0.32 (1)$& $8.60 (4)$ & $1.58 (12)$\\
   $7\cdots11.0$ & $24\cdots50$ & $1$ & $3$  & $0.26 (2) $  & $0.32 (1)$& $8.60 (5)$ & $1.78 (10)$\\
   $8\cdots9.25$ & $24\cdots50$ & $3$ & $1$  & $0.19 (13)$  & $0.32 (1)$& $8.61 (4)$ & $1.42 (17)$\\
   $8\cdots9.25$ & $24\cdots50$ & $1$ & $3$  & $0.36 (18)$  & $0.32 (2)$& $8.60 (4)$ & $1.45 (15)$\\
   $7\cdots11.0$ & $13\cdots50$ & $3$ & $3$  & $0.31 (1) $  & $0.33 (1)$& $8.70 (3)$ & $1.38 (10)$\\[1ex]\hline
\multicolumn{4}{|l||}{average:}
 & $0.28 (6)$ & $0.32 (3)$ & $8.66 (4)$ & $1.44 (13)$\\
\noalign{\smallskip}\hline
\end{tabular}
\end{table}
%\narrowtext
%%%%%%%%%%%%%%%%%%%%%%%%%%%%%%% END OF TABLE %%%%%%%%%%%%%%%%%%%%%%%%%%%%%%%%

%%%%%%%%%%%%%%%%%%%%%%%%%%%%%%% TABLE %%%%%%%%%%%%%%%%%%%%%%%%%%%%%%%%%%%%%%%
\clearpage\newpage
\begin{table}
\caption{\label{tab-chibar}
  Values of the critical disorder strength $W_{\rm c}$ and the critical
  exponent $\nu$ values computed from Equation (\ref{eq-polyfit}) with
  various $m$ values. The numbers in the $4$th and $5$th column denote
  orders $n_{\rm r}$ and $m_{\rm r}$ used in the expansions (10) and
  (11), respectively for which the best fits have been obtained. The
  goodness-of-fit $Q$ \cite{PreTVF92} is larger than $0.99$ in all
  cases.}
\begin{tabular}{|c|c|c|c|c|c|c|}
\hline\noalign{\smallskip}
%\hline
$m$  & $W$
     & $N$ & $n_{\rm r}$ & $m_{\rm r}$ &$W_{\rm c}$ &  $\nu$\\
\hline
2 & $7\cdots11.0$ &  $13\cdots50$  & $3$ & $1$  & $8.65 (2)$ & $1.36
(3)$\\ 2 & $7\cdots11.0$ &  $13\cdots50$  & $1$ & $3$  & $8.63 (2)$ &
$1.52 (4)$\\ 2 & $8\cdots9.25$ &  $24\cdots50$ & $3$ & $1$ & $8.57 (3)$
& $1.43 (6)$\\ 2 & $8\cdots9.25$ & $24\cdots50$ & $1$ & $3$ & $8.53 (3)$
& $1.65 (7)$\\[1ex]\hline
\multicolumn{5}{|l|}{average:}
&$8.60 (3)$ & $1.49 (5)$\\[1ex]\hline

3 & $7\cdots11.0$ &  $13\cdots50$  & $3$ & $1$  & $8.67 (2)$ & $1.29
(3)$\\ 3 & $7\cdots11.0$ &  $13\cdots50$  & $1$ & $3$  & $8.61 (1)$ &
$1.42 (3)$\\ 3 & $8\cdots9.25$ &  $24\cdots50$ & $3$ & $1$ & $8.55 (4)$
& $1.45 (8)$\\ 3 & $8\cdots9.25$ & $24\cdots50$ & $1$ & $3$ & $8.53 (4)$
& $1.51 (8)$\\[1ex]\hline
\multicolumn{5}{|l|}{average:}
&$8.59 (3)$ & $1.42 (6)$ \\[1ex]\hline

4 & $7\cdots11.0$ &  $13\cdots50$  & $3$ & $1$  & $8.54 (4)$ & $1.32
(9)$\\ 4 & $7\cdots11.0$ &  $13\cdots50$  & $1$ & $3$ & $8.51 (4)$ &
$1.40 (9)$\\ 4 & $8\cdots9.25$ &  $24\cdots50$ & $3$ & $1$ & $8.45 (10)$
& $1.56 (26)$\\ 4 & $8\cdots9.25$ & $24\cdots50$  & $1$ & $3$  & $8.42
(11)$ & $1.68 (28)$\\[1ex]\hline
\multicolumn{5}{|l|}{average:}
&$8.48 (7)$ & $1.49 (18)$\\[1ex]\hline\hline
\multicolumn{5}{|l|}{total average:}
&$8.56 (5)$ & $1.47 (10)$\\
%\hline
\noalign{\smallskip}\hline
\end{tabular}
\end{table}
%%%%%%%%%%%%%%%%%%%%%%%%%%%%%%% END OF TABLE %%%%%%%%%%%%%%%%%%%%%%%%%%%%%%%%

\end{document}